\documentclass[aps,reprint,prl]{revtex4-1}
\usepackage{amsmath}
\usepackage{amssymb}
\usepackage{graphicx}
\usepackage{hyperref}

\newcommand{\be}{\begin{equation}}
\newcommand{\ee}{\end{equation}}
\newcommand{\beq}{\begin{eqnarray}}
\newcommand{\eeq}{\end{eqnarray}}

\def\apj{{ApJ}}                 
\def\apjl{{ApJ}}                
\def\aap{{A\&A}}                

\def\mnras{{MNRAS}}             
\def\prc{{Phys.~Rev.~C}}        
\def\prd{{Phys.~Rev.~D}}        
\def\pasa{{PASA}}               
\def\ssr{{Space~Sci.~Rev.}}     
\def\nat{{Nature}}              

\begin{document}
\title{Are gravitational waves spinning down PSR J1023+0038?}
\author{B.~Haskell}
\affiliation{Nicolaus Copernicus Astronomical Center, Polish Academy of Sciences, ul. Bartycka 18, 00-716 Warsaw, Poland}
\author{ A.~Patruno}
\affiliation{Leiden Observatory, Leiden University, Neils Bohrweg 2, 2333 CA, Leiden, The Netherlands}
\affiliation{ASTRON, the Netherlands Institute for Radio Astronomy, Postbus 2, 7900 AA, Dwingeloo, the Netherlands}

\label{firstpage}

\begin{abstract}

The pulsar J1203+0038 rotates with a frequency $\nu\approx 592$ Hz and has been observed to transition between a radio state, during which it is visible as a millisecond radio pulsar, and and a Low Mass X-ray Binary state, during which accretion powered X-ray pulsations are visible. Timing during the two phases reveals that during the LMXB phase the neutron star is spinning down at a rate of $\dot{\nu}\approx -3 \times 10^{-15}$ Hz/s, which is approximately 27\% faster than the rate measured during the radio phase, $\dot{\nu}\approx -2.4 \times 10^{-15}$ Hz/s, and at odds with the predictions of accretion models. In this letter we suggest that the increase in spin-down rate  is compatible with gravitational wave emission, and in particular to the creation of a `mountain' during the accretion phase. We show that asymmetries in pycno-nuclear reaction rates in the crust can lead to a large enough mass quadrupole to explain the observed spin-down rate, which so far has no other self-consistent explanation, and that radio timing at the onset of the next millisecond radio pulsar phase can test this scenario. Another possibility is that an unstable $r$-mode with amplitude $\alpha\approx 5\times10^{-8}$ may be present in the system.

\end{abstract}


\maketitle



The system PSR J1023+0038 (from now on J1023) is a peculiar binary
that has been observed to transition back and forth between a radio
millisecond pulsar (RMSP) state and a low-mass X-ray binary (LMXB)
state~\cite{arc09}. The neutron star spins at a rate of ${\approx}592$
Hz and the companion star is a main sequence star of
${\approx}0.2\,M_{\odot}$.  Timing of the radio pulsations has led to
a precise measurement of the spin down of the pulsar
$\dot{\nu}=-2.3985\times 10^{-15}$ Hz/s.  After the last transition,
which occurred in June 2013~\cite{pat14,sta14}, X-ray observations
during the LMXB state of J1023 allowed to measure the accretion
powered pulsations~\cite{arc15} and the spin down of the accreting
pulsar $\dot{\nu}=-3.0413(90)\times 10^{15}$ Hz/s~\cite{jao16}, which
is approximately 27\% faster than the rate measured from timing during
the radio state.

The interpretation of the enhanced spin down as due to the interaction
between the accretion disk and the neutron star magnetosphere is
somewhat problematic because several inconsistencies remain in each
model considered (see~\cite{jao16} for an extended discussion). For
example, a propeller model (e.g.,~\cite{pap15}) with a standard
$\alpha-$disk where the inner portions are truncated and ejected from
the system, or an enhanced pulsar wind model~\cite{par16}, would require
a careful fine tuning of the model parameters to explain the close
match between the observed radio and LMXB spin down rates. The most
promising alternative scenario is the trapped-disk model
(see~\cite{spr93,dan10}) which was instead proposed to explain the
presence of outflows in the system and the peculiar low-luminosity of
J1023~\cite{jao16}. However, in this case the spin-down needs to
be enhanced both during the RMSP and the LMXB stage, meaning that no
difference in $\dot{\nu}$ should be observed.

Here we propose an alternative scenario that would solve this dilemma. We suggest that the additional spin-down is due to gravitational wave (GW) emission, triggered during the LMXB state.
Evidence for the presence of GWs in accreting neutron
stars is recently mounting due to the lack of sub-millisecond pulsars
(see e.g.~\cite{bha17}).  This may be due, in particular, to the
formation of `mountains', i.e. asymmetries in the mass distribution,
supported either by crustal or magnetic strains, or unstable modes of
oscillation \cite{Lasky15}. Note that these mechanisms have been considered before in
LMXBs \cite{bil98,and98,cut02,mel05}, but always in the context of
spin-balance, and it was generally found that it is not easy to build
a large enough mountain to balance the spin-up torque due to accretion
in these systems \cite{has11} (with the notable exception of some
persistently accreting systems where the mountain could potentially be
large enough for the neutron star to be spinning down during an
outburst), and that detection of these signals would require next
generation gravitational wave detectors, such as the Einsten
Telescope~\cite{wat08,has15}.

The situation here is, however, radically different. The accretion
rate is much lower (of the order of
$10^{-13}\,M_{\odot}\,$yr$^{-1}$~\cite{pap15,jao16}) and while, on the one
hand, this reduces the amount of accreted mass that can build the
`mountain', on the other it ensures that the spin-up torque is weak
enough to not contaminate the spin-down measurement. Furthermore the
precise spin-down rate obtained from radio timing allows for a
detailed comparison of the rates during the radio and accretion phase , which
is not possible for other LMXBs.

In conclusion the spin-down rate we attribute to GW emission is the
difference between the enhanced rate during the LMXB state and the
previous rate during the radio state, i.e.
\be
\dot{\nu}_{diff}=-6.428\times 10^{-16} \mbox{ Hz/s}.\label{add}
\ee
The spin-down rate due to GW emission is
\be
\dot{\nu}_{GW}\approx-1.4\times 10^{-13}\; \nu_{500}^5\;  I_{45}^{-1} \left(\frac{Q_{22}}{10^{37}\mbox{g cm$^2$}}\right)^2  \mbox{Hz/s},
\ee
where $Q_{22}$ is the mass quadrupole moment, $I_{45}$ the moment of
inertia of the neutron star in units of $10^{45}$ gm cm$^2$ and $\nu_{500}$ the spin frequency in units of 500 Hz. 
  
We can see that to explain the additional spin-down
in (\ref{add}) for the spin-frequency of J1023 one requires a
quadrupole of
\be Q_{22}=4.4\times 10^{35}\;
I_{45} \;\mbox{ g cm$^2$},
\ee
which corresponds to an ellipticity $\varepsilon\approx 5\times
10^{-10}$, well below the maximum that can be sustained
without breaking the crust, $\varepsilon_{MAX}\approx 10^{-5}$ ~\cite{Owen13}. Note that this is a conservative estimate of the GW contribution, as we have neglected the spin-up torque due to accretion which, although weak, may contribute to the spin-up at a level of $\dot{\nu}_{m}\approx10^{-16}\rm\,Hz\,s^{-1}$, for the maximum accretion rate of $\dot{M}=6\times10^{-13}\rm\,M_{\odot}\,yr^{-1}$.
Let us thus consider some of the most likely models to establish whether they may lead to such a quadrupole in J1023.


First of all we will consider the scenario in which asymmetries in the local accretion rate and crustal composition can lead to asymmetric heat release due to pycnonuclear reactions in the crust, i.e. `deep crustal heating' \cite{HZ90}, that will source a mass quadrupole~\cite{bil98,ush00}:
\be
Q_{22}=3\times 10^{35} R_{12}^4\left(\frac{\delta T_q}{10^5}\right)\left(\frac{E_{th}}{30\;\mbox{MeV}}\right)^3\mbox{g cm$^2$},
\ee
where $R_{12}$ is the radius in units of 12 km, $E_{th}$ is the threshold energy for the pycnonuclear reactions responsible for deep crustal heating and $T_q$ is the quadrupolar temperature increase due to the reactions (which will only be a fraction of the total heating $\delta T$). Rearranging we see that we require a quadrupolar temperature increase of
\be
\delta T_q\approx 1.5 \times 10^{5} \;R_{12}^{-4}\;I_{45}\; \left(\frac{E_{th}}{30\;\mbox{MeV}}\right)^{-3}\;\mbox{ K}.\label{quadt}
\ee
Is such a quadrupolar temperature increase possible in J1023? The total local increase in temperature due to pycnonuclear reactions is~\cite{ush01}:
\be
\delta T\approx 10^{2} \;C_k^{-1}p_{30}^{-1} Q_M \Delta M_{21}\;\mbox{ K},\label{temp}
\ee
where $C_k$ is the heat capacity in units of the Boltzman constant per
baryon, $p_{30}$ is the pressure in units of $10^{30}$ erg/cm$^3$,
$Q_M$ is the heat released locally by the reactions per accreted baryon,
in units of MeV, and $M_{21}$ is the accreted mass in units of $10^{21}$ g. To obtain
an estimate from the above expression we will take an accretion rate
of $5\times 10^{-14}
M_\odot/\mbox{yr}\lesssim \dot{M} \lesssim 6 \times 10^{-13} M_\odot$/yr (estimated by~\cite{pap15}) and
thus consider that in a year of accretion the system can
accrete $\Delta M\approx 10^{21}$ g.

To obtain the heat capacity we first need to estimate the temperature
of the neutron star which is currently unconstrained from X-ray observations (which are
dominated by the thermal emission of the hot polar caps during the RMSP state and by the accretion
induced X-ray radiation during the LMXB state~\cite{bog11,bog15}). To do
this let us consider heating due to deep crustal heating at a rate~\cite{bro98}:
\be
W_{CH} = 6\times 10^{30} \left(\frac{\dot{M}}{10^{-13} M_\odot/yr}\right)\;\mbox{erg/s},
\ee
which will be balanced by photon cooling at the surface
\be
L_{ph}=1.7\times 10^{33} R_{12}^2 \left(\frac{T_s}{10^{6} K}\right)^4 \;\mbox{erg/s},
\ee
with $T_s$ the surface temperature, which for an iron envelope can be related to the core temperature $T$ by the relation~\cite{gud83}
\be
\left(\frac{T_s}{10^{6} \mbox{K}}\right)^4=2.42\; g_{14}\; \left(18.1\frac{T}{10^{9} K}\right)^{2.42},
\ee
with $g_{14}$ the surface gravity in units of $10^{14}$ cm/s$^2$, or by Urca reactions if the star is massive enough, at a rate
\be
L_{Urca}=10^{33} \left(\frac{T}{2\times 10^{7} K}\right)^6 \left(\frac{R_c}{3\mbox{km}}\right)^3\;\mbox{erg/s},
\ee
with $R_c$ the radius of the core region in which Urca reactions can proceed. For both cooling mechanisms, and taking the maximum estimated accretion rate during outburst, we obtain $T\lesssim 10^{7}$ K for the star.
At these temperatures the heat capacity per baryon in units of the Boltzman constant is~\cite{pot15} $C_k \approx 10^{-6}$ at $\rho\approx 10^{12}$ g/cm$^3$, which is approximately the density close to the neutron drip point,  at which most of the heating occurs ( with $E_{th}=30$ MeV, $Q_M=0.5$ MeV and $p_{30}=1$).

From (\ref{temp}) we obtain a total heating rate of $\delta T \approx 5\times 10^{6}$ K for an accreted
mass of $\Delta M=10^{20}$ g, which is what J1023 is expected to have
accreted on the order of a month during the LMXB state. 

In order to build a large enough quadrupole we see from (\ref{quadt}) that we would need (although note that deeper layers will also contribute to the quadrupole, thus reducing the required heating in a single layer at neutron drip):
\be
\frac{\delta T_{q}}{\delta T}\gtrsim 3 \times 10^{-2}.
\ee

There is no firm estimate of this quantity, with the only limits coming from the non-detection in X-rays of quadrupolar flux perturbations in quiescence in transiently accreting LMXBs \cite{ush00, has15} which sets ${\delta T_{q}}/{\delta T}\lesssim 0.1$.

Furthermore we may expect asymmetries in the accreted mass at the surface to be confined on a Rossby adjustment radius \cite{SLU02}, $R_a=\sqrt{(p/\rho)}/4\pi\nu\approx 3\times 10^5$ cm for J1023 with $p=10^{30}$ erg/cm$^3$ and $\rho=10^{12}$ g/cm$^3$. The rapid rotation rate of the source may thus allow for asymmetries in composition imprinted by accretion at the surface to persist also deep in the crust.

In conclusion it is likely that a large enough quadrupole can be built
on J1023 to explain the additional spin-down. After the accretion
phase is over the mountain will be washed away on a
thermal timescale for the crust \cite{bro98} $\tau_{th}\approx 0.2 \;
p_{30}^{3/4}$ yrs, although note that deeper layers, at higher
pressures than the ones we consider, may also contribute to the
quadrupole and thus dissipate on longer timescales.

Nevertheless compositional asymmetries may be
frozen in~\cite{ush00} and would allow to 'build' the mountain over
successive accretion phases. If this is the case we would predict the
increase in spin-down rate to remain even after the LMXB state and for
the measured value in radio during the next quiescent state to be the
same as the current rate in X-rays.


For mountains sustained by magnetic stresses one has~\cite{Shib89}
\be
Q_{22}\approx 5 \times 10^{32} \Delta M_{21}\; \mathcal{A}\; \left(1+\frac{\Delta M}{M_c}\right)^{-1}\mbox{g cm$^2$},\label{quadro}
\ee

where $\mathcal{A}$ is a constant of order unity that depends on the
equation of state ~\cite{mp05} and $M_c\approx 10^{-7} (B/10^{12} G)^{4/3} M_\odot$
is the critical mass at which the amplitude of the quadrupole saturates.
Note that close to the critical mass the simple estimate in (\ref{quadro}) is no
longer accurate and numerical simulations are necessary ~\cite{pay04}.
In general close to the critical mass one finds that the external
dipolar magnetic field is reduced by approximately an order of
magnitude by field burial, although numerical simulations seem to
indicate that while the quadrupole saturates, magnetic burial does not,
and may reduce the field even further~\cite{VM09, Priymak11}.
Despite the uncertainties, the estimate in (\ref{quadro}) suggests that a large enough magnetic mountain cannot be built on J1023 during an accretion phase, as the required amount of mass would take much longer to be accreted. We will thus not consider this mechanism further.

Another possibility is that modes of oscillation of the star may grow unstable during the accretion phase, and provide the additional gravitational wave spin-down torque. The main candidate for this mechanism is the $r$-mode \cite{and98}, as the $f$ mode instability will be stabilised by superfluid mutual friction for temperatures below $\approx 10^9$ K \cite{aghdamp}.
For an internal temperature of $T\approx 10^7$ K and $\nu\approx 592$ Hz, standard models of hadronic neutron stars would predict J0123 to be r-mode unstable (although see~\cite{HO11,has12,gus16} for a discussion of why additional physics is probably required in these models).
The spin-down rate due to an unstable r-mode of dimensionless amplitude $\alpha$ is, if we assume an $n=1$ polytrope for the equation of state, \cite{Owen98}:
\be
\dot{\nu}\approx 6.7\times 10^{-16} \left(\frac{\alpha}{10^{-7}}\right)^2 M_{1.4} R_{12}^4 \;\nu_{500}^7\;\mbox{Hz/s}.
\ee
where $M_{1.4}$ is the neutron star mass in units of 1.4$M_{\odot}$.
For our source we thus require
\be
\alpha\approx 5.5 \times 10^{-8} M_{1.4}^{-1/2} R_{12}^{-2},
\ee
which is well below theoretical estimates of saturation amplitudes~\cite{has15b} and consistent with observational upper limits on r-mode amplitudes in LMXBs~\cite{has12, mah13, sch17}.
It is also well below current upper limits set by LIGO \cite{LIGO15}.

We can also estimate the heating that the r-mode would produce
\be
W_r \approx 4.5\times 10^{33} \left(\frac{\alpha}{10^{-7}}\right)^2 M_{1.4}^2\; R_{12}^6\; \nu_{500}^8\;\mbox{erg/s},
\ee
which, balanced by direct Urca reaction gives $T\approx 2\times 10^7$ K thus potentially contributing to reheating the system more than deep crustal heating.

It is thus possible that the system lies close to the instability
curve, and is pushed into the unstable region by
heating due to deep crustal reaction. The r-mode can then grow unstable
and contribute to the observed spin-down increase, heating the system further. This is an
interesting possibility, as it would indicate that the saturation
amplitude of the mode is indeed small, of the order of $\alpha\approx
10^{-8}-10^{-7}$, which challenges most theoretical models and requires additional physics to be included in the picture, such as, for example,
the existence of a phase transition to quark matter in the core
\cite{alf15}.


There are two possible observational tests that can be performed to
verify whether GWs are the main cause of the
excess spin down in J1023. The first relies on timing the pulsations
during the RMSP state. In this case if the quadrupolar asymmetry
generating the GWs is dissipated on a specific
timescale, then the excess spin down should be observed to disappear
on the same timescale. According to our estimates the main contribution to the quadrupole is from layers close to neutron drip and will dissipate on a thermal timescale of a few months, with contributions from deeper layers dissipating on longer timescales of a few years. If the excess spin down is instead the result
of the interaction of the neutron star magnetosphere with the
accretion disk then the excess spin down should disappear sharply once
the transition to the RMSP state is completed.  On the other hand if
the mountain is being built cumulatively over successive LMXB states
as may be the case if compositional asymmetries are frozen into the crust, then the enhanced spin down will persist during the next RMSP state.

The second test can be performed during the LMXB state and involves
prolonged timing of the X-ray pulsations observed during the accretion
process. In this case if the `mountain' builds up over time as additional mass is accreted, the enhanced spin down should be observed to
increase approximately linearly over time, until the mechanism saturates (assuming the
mass accretion rate remains relatively constant, which is a very
plausible hypothesis in J1023~\cite{pat14,bog15}).

Furthermore, if in the future the surface temperature of the neutron star were to be measured, and resulted in an estimate of the core temperature of $T>10^7$ K, this would suggest that an additional heating mechanism, in addition to deep crustal heating, is active, supporting the hypothesis that the r-mode instability is active. Confirming the existence of an unstable r-mode in J1023 would allow us to constrain the instability window and the saturation amplitude of the mode, thus constraining the interior physics of neutron stars \cite{has15, KS16}.

Finally we note that, for a distance to the source of $1.4$
kpc~\cite{del12}, the measured gravitational wave strain would be
$h_0\approx 6\times 10^{-28}$, which is below the detection limit for
current interferometers, but is potentially detectable by next
generation interferometers such as the Einstein Telescope, if the
signal is long lived and can be integrated over outburst timescales of
the order of a few years. If thermal and compositional asymmetries, such as those
calculated here, are typical for LMXBs, however, other rapidly
rotating sources with higher accretion rates are likely to be good
targets for current GW searches \cite{has15}.

This project has received funding from the European Union's Horizon 2020 research and innovation programme under grant agreement No. 702713. AP acknowledges support from an Netherlands Organization for Scientific Research  (NWO) Vidi grant.



\bibliography{biblio}


%
\label{lastpage}

\end{document}